\documentclass[12pt]{llncs}
\usepackage[letterpaper,hmargin=1.25in,vmargin=1.25in]{geometry}
\usepackage{graphics}
\usepackage{graphicx}
\usepackage[english]{babel}
\usepackage{amsmath}
\usepackage{amssymb}
\usepackage{amsfonts}

\makeatletter                     % https://tex.stackexchange.com/questions/179997/using-package-appendix-with-llncs
\newcommand{\@chapapp}{\relax}%
\makeatother

\usepackage[title]{appendix}
\usepackage{booktabs}

\usepackage[linesnumbered,ruled,vlined]{algorithm2e}

\usepackage{latexsym}
\usepackage{epsfig}
\usepackage{hyperref}
\hypersetup{
    pdfcreator={},
    pdfproducer={}
}
\usepackage[latin1]{inputenc}
\usepackage{times}
\usepackage[T1]{fontenc}

\usepackage{url}
\usepackage{lscape}

\title{Computing the Weight Distribution of the Binary Reed-Muller Code ${\mathcal R} (4,9)$}

\author{Miroslav Markov  and
Yuri Borissov}

\institute{
  }

\begin{document}

\maketitle
\pagestyle{plain}

\begin{abstract}\noindent
We compute the weight distribution of the ${\mathcal R} (4,9)$ by combining the approach described in D.~V.~Sarwate's Ph.D. thesis from 1973 with knowledge on the affine equivalence classification of Boolean functions. To solve this problem posed, e.g., in the MacWilliams and Sloane book \cite[p. 447]{McWSlo}, we apply a refined approach based on the classification of Boolean quartic forms in $8$ variables due to Ph.~Langevin and G. Leander, and  recent results on the classification of the quotient space ${\mathcal R} (4,7)/{\mathcal R} (2,7)$ due to V.~Gillot and Ph.~Langevin. %The work is still in progress.
\end{abstract}
\vspace*{-0.5cm}
{\bf keywords:} code weight distribution; binary Reed-Muller code

\section{\bf Introduction}
\label{sect0}
\medskip
For basic coding theoretical notions, we refer to \cite{McWSlo}.

It is well known that the weight enumerator of a linear code enables one to determine the probabilities of decoding error and decoding failure when that code is utilised to transmit information across a noisy binary symmetric channel (BSC) and is decoded by a maximum-likelihood bounded-distance decoding algorithm.
%in order to determine the probability of decoding error or the probability of decoder failure one must know the weight distribution (enumerator) of the utilized code.

The binary Reed-Muller codes form one of the oldest studied families of codes invented in 1950s and have an easy to implement decoding algorithm based on majority-logic circuits. However, there are few general results about their weight structure. Namely, the weight distribution of those codes is known only for:
         \begin{itemize}
          \item the 1st and 2nd-order \cite{SloBer} (1970);
          \item arbitrary order when the weight $< 2d_{min}$ \cite{KasTok} (1970), and has been extended for weights $< 2.5d_{min}$ in \cite{KasTokAzu} (1976);
%          \item weight divisibility: the McEliece theorem \cite{McE}.
         \end{itemize}

For information about the weight distributions of binary Reed-Muller codes of particular orders and lengths, the reader is directed to \cite{OEIS}.
In particular, it is worth pointing out four works concerning the third and fourth order Reed-Muller codes \cite{KasTokAzu}, \cite{SugKasFuj} - \cite{vTi}, as well as, the very recent work on the weight spectrum of some families of  binary Reed-Muller codes \cite{CarSol}.

This paper is organized as follows. In the next section we give some necessary preliminaries. In Section \ref{sect3}  a refined approach to the problem under consideration allowing to save computational efforts is exposed. %Section \ref{sect3} provides an application of that approach to the  binary Reed-Muller codes ${\mathcal R} (4,8)$ and  ${\mathcal R} (4,9)$.
Some conclusions are drawn in the last section.

\section{ Preliminaries}
\label{sect2}
\medskip

For basic knowledge on Boolean functions and their applications in cryptography and coding theory, we direct the reader to \cite{Car} and \cite{CusSta}.
Herein, for the sake of completeness, we recall the classical definition of the binary Reed-Muller code.
\begin{definition}
The $r-$th order binary Reed-Muller (or RM) code ${\mathcal R}(r,m)$ of length $n = 2^{m}$, for $0 \leq r \leq m,$ is the set of all binary vectors $\bf{f}$ of length $n$ which are truth tables of Boolean functions $f({\bf x}), {\bf x} = (x_{1},\ldots,x_{m})$, having algebraic normal forms of degree at most $r$.
\end{definition}
%In this paper, we consider only binary RM codes, i.e., over the alphabet $\mathbb{F}_{2}$.
\noindent Henceforth the binary vector $\bf{f}$ of length $2^{m}$ will be identified with corresponding Boolean function $f$ in $m$ variables.% (which is a slight abuse in notations).

In order to present our results we need to remind the definitions of weight distribution/enumerator of a code, i.e., an arbitrary set $\bf C$ of vectors with fixed length $n$. (All considered codes in this paper are binary, and these definitions hold in particular for cosets of binary linear codes.)

\begin{definition}
The weight distribution of a code ${\bf C}$ of length $n$ is the vector $W({\bf C}) = (W_{0}, \ldots , W_{n})$, where $W_{i}$
denotes the number of words with Hamming weight $i$.
\end{definition}

\begin{definition}
Weight enumerator of a code ${\bf C}$ with  weight distribution $W({\bf C}) = (W_{0}, \ldots , W_{n})$ is defined as the following polynomial in the indeterminate $z$:
${\mathcal W}[z;{\bf C}] = \sum_{i = 0}^{n} W_{i}z^{i}$.
% Sometimes it is more convenient to use the polynomial in two indeterminates $x,y$: \hspace*{4.5cm} ${\mathcal A}(x,y;{\bf C}) = \sum_{i = 0}^{n} A_{i}x^{i}y^{n-i}.$%, e.g. the MacWilliams identities for linear codes. \end{justify}
\end{definition}
%Note that, in particular, these definitions are valid for the cosets of binary linear codes.

In this paper, we make use of two facts first time exposed in \cite{DVS}, and adapted in the next two theorems. For $0 \leq r \leq m,$ we will denote by ${\mathcal H}^{(r)}(m)$ the set of all homogeneous polynomials on $m$ binary variables of algebraic degree $r$ adjoined with the $0$.
\medskip

\begin{theorem}\label{th.1}(\cite[5.12]{DVS})
For $0 \leq r \leq m,$ it holds:
\[
{\mathcal W}[z;{\mathcal R}(r+2,m+2)] = \hspace{-0.5cm}\sum_{p \in {\mathcal H}^{(r+2)}(m+1)} {\mathcal W}^{2}[z;p+{\mathcal R}(r+1,m+1)].
\]
\end{theorem}
\medskip

\begin{theorem}\label{th.2}(\cite[5.13]{DVS})
Let $p =  e + fx_{m+1}$, with given $e \in {\mathcal H}^{(r+2)}(m)$ and $f \in {\mathcal H}^{(r+1)}(m)$.
Then the weight enumerator of the coset ${\mathcal C}(p) = p + {\mathcal R}(r+1,m+1)$ equals to:
\[
 \hspace{-0.05cm}(*)\;\; \sum_{g \in {\mathcal H}^{(r+1)}(m)}  {\mathcal W}[z;e+g+{\mathcal R}(r,m)] \cdot {\mathcal W}[z;e+g+f+{\mathcal R}(r,m)].
\]
\end{theorem}

For definition of the affine general linear group ${GA}(m)$ and its subgroup the general linear group ${GL}(m,2)$, we refer to \cite[Ch.13.9]{McWSlo}.
The action of $A \in {GA}(m)$ on a Boolean function $f(\bf{x})$ is denoted by $f \circ A$, i.e., $f \circ A = f(A{\bf x})$.\\
Another necessary definition is that of affine equivalence of two cosets of Reed-Muller code:
\begin{definition}
The cosets $C_{1}$ and $C_{2}$ of ${\mathcal R}(r,m)$ %such that $C_{1} = f_{1} + {\mathcal R}(r,m)$ and $C_{2} = f_{2} + {\mathcal R}(r,m)$
are called affine equivalent if there exist $f_{1} \in C_{1}, f_{2} \in C_{2}$ and a transformation $A \in {GA}(m)$ such that
$f_{1} \circ A = f_{2}$.
\end{definition}
%\noindent If the cosets $f_{1} + {\mathcal R}(r,m)$ and $f_{2} + {\mathcal R}(r,m)$ are affine equivalent, we will write $f_{1} \equiv f_{2} \pmod{{\mathcal R}(r,m)}$.
\medskip

\noindent
%If such a $A$ exists, then it is easy to see that under the action of $A$, the coset $C_{1}$ is mapped onto the coset $C_{2}$.
In this article, we extensively use the following evident property of the affine equivalent cosets:\\
\noindent
${\mathcal P:}$ {\it The weight distributions/enumerators of two affine equivalent cosets of any Reed-Muller code coincide.}
\medskip

Affine equivalence classification of the cosets of RM codes is useful in studying important coding theoretical and cryptographic properties of Boolean functions comprising them. In work \cite{LanLea}, the authors have present a strategy used to compute the complete classification of Boolean quartic forms in eight variables, i.e., the classification of the quotient space ${\mathcal R}(4,8)/{\mathcal R}(3,8)$ under the action of ${GL}(8,2)$. Here, just as an extract of this result, we point out that the Boolean quartic forms of eight variables can be classified in $999$ (see, also \cite{Hou}) linear equivalence classes listed in \cite{Lan}. In fact, from information presented in \cite{Lan}, we make use essentially only of the representative and orbit size of each class. The authors of \cite{LanLea} have also outlined some applications of their result including a refined estimation of the covering radius of ${\mathcal R}(3,8)$ and counting the bent functions in $8$ variables. % In this paper, we point out yet another one, i.e., computing the weight distribution of ${\mathcal R}(4,9)$.\\
Recently, an interest in that topic has been renewed by the work \cite{GilLan} which (among other things) provides a classification
${\mathcal R}(4,7)/{\mathcal R}(2,7)$, and deduces some consequences on the covering radius of ${\mathcal R}(3,7)$ and the classification of near bent functions. In Section \ref{sect3}, we shall demonstrate how to use the two classifications aforesaid to compute the weight distribution of the ${\mathcal R}(4,9)$.

\section{ The refined approach}
\label{sect3}
\medskip

%\label{sect2}

\subsection{Rationale}
Now, we describe a strategy following which makes feasible the computation of ${\mathcal W}[z;{\mathcal R}(4,9)]$. %Since Theorems \ref{th.1} and \ref{th.2} are stated in general terms our explanations will keep this manner.

First, to apply efficiently Theorem \ref{th.1} we may use the linear equivalence classification of the quotient space ${\mathcal R}(r+2,m+1)/{\mathcal R}(r+1,m+1)$ as follows. Assume $n(r+2,m+1)$ is the class number, and let %, i.e. the classification of Boolean $(r+2)-$order forms of $m+1$ variables.
a representative $p_i \in {\mathcal H}^{(r+2)}(m+1)$ and the orbit size $L_{i}$ of the $i-$th class are known where $1 \leq i \leq n(r+2,m+1)$.
Then the formula in Theorem \ref{th.1} can be rewritten as:
\begin{eqnarray}\label{f.1}
{\mathcal W}[z;{\mathcal R}(r+2,m+2)] = \hspace{-0.5cm}\sum_{i = 1}^{n(r+2,m+1)} L_{i} {\mathcal W}^{2}[z;p_{i}+{\mathcal R}(r+1,m+1)].
\end{eqnarray}
So, this technique reduces the number of the necessary weight distribution computations to the class number $n(r+2,m+1)$ which is significantly smaller than $2^{\binom{m+1}{r+2}} = \vert{\mathcal H}^{(r+2)}(m+1)\vert$. For instance, as it has been already mentioned, $n(4,8) = 999$ which should be compared with $2^{70}$.

Second, to compute more efficiently the weight enumerator of the coset ${\mathcal C}(p) = p + {\mathcal R}(r+1,m+1)$ by expression (*) in Theorem \ref{th.2}, we can use the affine equivalence classification of ${\mathcal R}(r+2,m)/{\mathcal R}(r,m)$. To this end, let us recall the following definition:
\begin{definition}
The subgroup ${\mathcal St}(e)$ of ${GA}(m)$ that fixes $e \in {\mathcal H}^{(r+2)}(m)$, i.e. for each $A \in {\mathcal St}(e)$ it holds: $e \circ A \in  e + {\mathcal R}(r+1,m),$ is called stabilizer of $e$ in ${GA}(m)$.
\end{definition}
%Henceforth it is assumed a fixed $e \in {\mathcal H}^{(r+2)}(m)$.
For  given $e \in {\mathcal H}^{(r+2)}(m)$, the stabilizer ${\mathcal St}(e)$ partitions the cosets of form $e+g+{\mathcal R}(r,m), g \in {\mathcal H}^{(r+1)}(m)$ into disjoint orbits. Let us denote by $\Delta(e)$ this partition. Then some considerations helping to reduce the cost of needed computations can be summarized by the next rationale:
\medskip

${\bf{ \mathcal F}:}$ {\it As it follows by Property ${\mathcal P}$, the enumerator ${\mathcal W}[z;e+g+{\mathcal R}(r,m)]$ is preserved when $g$ runs over an orbit of $\Delta(e)$. This permits the computation of the expression (*) in  Theorem \ref{th.2} to be performed more efficiently. Namely, putting the common enumerator "outside of brackets", the number of needed polynomial multiplications for computing ${\mathcal W}[z;{\mathcal C}(p)]$ will be equal to the number of orbits in $\Delta(e)$ while that of polynomial additions is retained to (almost) $2^{\binom{m}{r+1}}$, of course. Besides, if for some orbits those weight enumerators are identical, this trick could be extended to reduce an extra number of multiplications. In summary, the number of necessary polynomial multiplications is equal to the number $s = s(e)$ of distinct weight enumerators ${\mathcal W}[z;e+g+{\mathcal R}(r,m)]$ when $g$ runs over the set of representatives of $\Delta(e)$'s orbits.}\\
As we shall show in the next subsection, the above described phenomenon is very common when $e \in  {\mathcal H}^{(4)}(7)$.
\medskip

\subsection{Computing ${\mathcal W}[z;{\mathcal R}(4,9)]$}

Our computational work is divided into two main phases: a pre-computing and an actual computing.
\medskip

The aim of pre-computing is to provide tools for efficient computation of the expression (*) in Theorem \ref{th.2} given a specific $e$ and $f$.
Let ${\mathcal E}(4,7)$ be the set of representatives of the twelve linear equivalence classes of ${\mathcal R}(4,7)/{\mathcal R}(3,7)$ given in \cite{Lan1}.
\medskip

\noindent For fixed $e \in {\mathcal E}(4,7)$, the pre-computing involves the following three tasks:
\begin{itemize}
\item ${\mathcal T1:}$ Constitute and store the orbits of the partition $\Delta(e)$;
\item ${\mathcal T2:}$ Compute the weight enumerators of the cosets $e+g+{\mathcal R}(2,7)$ when $g$ varies over a set of representatives of $\Delta(e)$'s orbits;
\item ${\mathcal T3:}$ Merge the orbits with identical weight enumerators (obtaining a coarse partition $\Delta^{\prime}(e)$)  and make data arrangement permitting for given $f \in  {\mathcal H}^{(3)}(7)$ to look up the identifier of a block in the $\Delta^{\prime}(e)$ containing $e+f+{\mathcal R}(2,7)$ (respectively, to have direct access to the common weight enumerator).
\end{itemize}
\noindent
For all $e \in  {\mathcal E}(4,7)$, we present in {\bf Table 1.} of the {\bf Appendix A} the sizes |$\Delta(e)$| and |$\Delta^{\prime}(e)$| of partitions $\Delta(e)$ and $\Delta^{\prime}(e)$, respectively.
\begin{remark} We would like to note that:
\begin{itemize}
\item The task ${\mathcal T1}$ is efficiently performed based on the so-called "orbit algorithm" \cite{AHu} using the set of  generators of the stabilizer ${\mathcal St}(e)$ provided by \cite{Lan1};
\item The task ${\mathcal T2}$ can be carried out simultaneously for all representatives by exhaustive generation of the codewords of ${\mathcal R}(2,7)$ based on some Gray code.
\end{itemize}
\end{remark}
\medskip

Now, following ${\mathcal F}$ and Theorem \ref{th.2}, we present an algorithm for computing the weight enumerator ${\mathcal W}[z ; C(p)]$ of the coset $C(p) = p + {\mathcal R}(3,8)$ where $p = e + fx_{8}$ for fixed $e \in {\mathcal E}(4,7)$ and a given input $f \in {\mathcal H}^{(3)}(7)$.
Note that it can be implemented as a subroutine. Recall also that the common weight enumerator $w_{i}(z)$ corresponding to the block $\delta_{i}$ in $\Delta^{\prime}(e)$  has been  already computed in the pre-computing task ${\mathcal T2}$ where $1 \leq i \leq |\Delta^{\prime}(e)| = s(e)$. % denotes the size of $\Delta^{\prime}(e)$, i.e., the number of those blocks.

\vspace{-0.40cm}
\begin{algorithm}[htbp]\label{alg.1}
\caption{\small{Returning the weight enumerator ${\mathcal W}[z ;  C(p)]$ where $p = e + fx_{8}$ for fixed $e$ and a given $f \in {\mathcal H}^{(3)}(7)$}}\vspace{0.20cm}
\footnotesize{
U[z] := 0\;
\For {i in [1,$s(e)$]} {
    UU(z) := 0\;
    \For {g in $\delta$[i]} {
        j := FindBlock(g+f)\;
        UU(z) := UU(z) + w[j](z);
    }
    U(z) := U(z) + w[i](z) * UU(z);
}
W[z; $C(p)$] := U(z);
}
\end{algorithm}

\vspace{-0.40cm}
In the actual computing, we apply formula (\ref{f.1}) supposing a set ${\mathcal S}$ of pairs: (representative $p_{i}$, orbit size $L_{i}$) for the $i-$th class $O_{i}, 1 \leq i \leq 999$, of the classification of ${\mathcal R}(4,8)/{\mathcal R}(3,8)$ is available.
W.l.o.g., we may assume each $p_{i}$ is of the form $e + f_{i} x_{8}$ for some $e \in {\mathcal E}(4,7)$ and $f_{i} \in {\mathcal H}^{(3)}(7)$, so the set of classes is naturally partitioned into subsets ${\mathcal O}(e)$ of cardinalities $n(e), e\in {\mathcal E}(4,7)$.
(The values $n(e)$ are given in the first column of {\bf Table 2.} of the {\bf Appendix A}.) Evidently, we have $\sum_{e \in {\mathcal E}(4,7)}n(e) = 999$.
Bellow, we present an algorithm for computing the sum in formula (\ref{f.1}) and thus ${\mathcal W}[z;{\mathcal R}(4,9)]$. Note that we call the subroutine computing ${\mathcal W}[z ; C(p)]$.% requested by Theorem \ref{th.1} for fixed $e \in {\mathcal E}(4,7)$.

\vspace{0.30cm}
\begin{algorithm}[H]\label{alg.2}
\caption{\small{Computing ${\mathcal W}[z;{\mathcal R}(4,9)]$}}\vspace{0.20cm}
\footnotesize{
V(z) := 0\;
\For {$e \in {\mathcal E}(4,7)$} {
    \For {j in [1,$n(e)$]} {
        p := representative(${\mathcal O}(e)$[j])\;
        L := size(${\mathcal O}(e)$[j])\;
        V(z) := V(z) + L * ${\mathcal W}^{2}$[z;C(p)];
    }
}
${\mathcal W}[z;{\mathcal R}(4,9)] = V(z)$;
}
\end{algorithm}

The data present in \cite{Lan} contains information to form a set ${\mathcal S}^{\prime}$ of kind similar to ${\mathcal S}$. However, the representatives $p^{\prime}_{i}$ there are of the form $e^{\prime} + f^{\prime}_{i}x_{8}$ where $e^{\prime}$s constitute different set of representatives of the twelve classes of ${\mathcal R}(4,7)/{\mathcal R}(3,7)$, say ${\mathcal E}^{\prime}(4,7)$. For some elements of ${\mathcal E}(4,7)$ and ${\mathcal E}^{\prime}(4,7)$, their linear equivalence is evident by eye inspection. For the remaining, we determined those which are linearly equivalent by computing the vectors of invariants of their duals (see, for details \cite[pp. 115-117]{Hou}). The matching found is represented in the rows of {\bf Table 2.} where ${\overline{\mathcal E}(4,7)}$ and ${\overline{\mathcal E}^{\prime}(4,7)}$ are the sets consisting of dual forms of those in ${\mathcal E}(4,7)$ and ${\mathcal E}^{\prime}(4,7)$, respectively. To find out a nonsingular $(7 \times 7)$ matrix ${\bf A}$ with property that $e^{\prime} \circ {\bf A} \in e + {\mathcal R}(3,7)$ for thus determined pairs $(e^{\prime}, e)$, we wrote a simple program in {\it C} which generates at random such a nonsingular square matrix and then checks the imposed condition. This technique is sufficiently effective (due to relatively large stabilizers sizes, see, \cite[{\bf Table 2.}]{LanLea}) and the program finished successfully its work in reasonable time. The results are presented in the last column of {\bf Table 2.} of the  {\bf Appendix A}. Finally, acting on corresponding  $f^{\prime}_{i}, 1 \leq i \leq 999$ by the linear transformations got (of course, ignoring the terms of degree less than $3$), we are yielded with type of a set requested by the {\bf Algorithm 2}. The obtained weight distribution is presented in the {\bf Appendix B}.

\subsection{Evaluating the computational costs}
For details about computational costs of task ${\mathcal T1}$ of the pre-computing, we refer to \cite{GilLan} and \cite{AHu}. The computational complexity of task ${\mathcal T2}$ is in total proportional to the product $68443 \times 2^{29} \approx 2^{45.06}$ with the first factor being the number of classes of ${\mathcal R}(4,7)/{\mathcal R}(2,7)$  and the second being the size of ${\mathcal R}(2,7)$. Task ${\mathcal T3}$ can be carried out by applying some sorting technique. In summary, the pre-computing in case $r = 2$ and $m = 7$ is efficiently performed. In addition, we note that the compressed storing of orbit and data arrangement into RAM needs at most $124$ GB of memory.

In the actual computing, for every $e \in {\mathcal E}(4,7)$, {\bf Algorithm 1} requires $|\Delta^{\prime}(e)|$ multiplications and about $2^{35}$ additions of degree $128$ polynomials with nonnegative integer coefficients. Therefore, {\bf Algorithm 2} requires $\sum_{e \in {\mathcal E}(4,7)}n(e) \times |\Delta^{\prime}(e)| = 1 827 252 \approx 2^{20.8}$ multiplications and about $999 \times 2^{35} \approx 2^{45}$ additions of polynomials of that kind, and $999$ squarings of degree $256$ polynomials, of course.

%%%%%%%%%%%%%%%%%%%%%%%%%%%%%%%%%%%%%%%%%%
\section{Conclusion}
\label{sect6}

In concluding remarks of his Ph.D. thesis \cite{DVS}, Dilip~V.~Sarwate has discussed the applicability of methods developed there to longer Reed-Muller codes, say of lengths $512$ and above. He has estimated and come into conclusion that there are too many equivalence classes of cosets of the ${\mathcal R}(2,7)$ in  ${\mathcal R}(4,7)$ in order to be useful in enumerating the ${\mathcal R}(4,9)$. However, as it is shown in this paper, due to the recent advancements in classification of Boolean functions \cite{GilLan},\cite{LanLea} and utilization of modern powerful computers, the solution of that long-standing problem is obtained successfully. Nevertheless, it seems likely that the method has almost reached its limits of utility as far as further enumerations are considered.
\medskip

\noindent
P.S. After submitting the first version of this paper, we saw on Philippe Langevin's numerical project page that the classification of Boolean cubic forms in $9$ variables allowed them (together with Eric Brier) to compute the weight distribution of the Reed-Muller codes of order $3$ in $9$ and $10$ variables. However, the details of their method are not exhibited. Finally, we would like to note that a sort of a refined approach as this one presented in our paper can be also applicable to these latter codes.

%\end{landscape}

\section*{\it \bf Acknowledgments} We would like to thank Vladimir D. Tonchev for his stimulating discussions on the topic considered and providing the copy of Ph.D. thesis \cite{DVS}.
This work was supported, in part, by the Ministry of Education and Science of Bulgaria under the Grant No. DO1-168/28.07.2022 "National Centre for High-performance and Distributed Computing" (NCHDC). The authors acknowledge the provided access to the e-infrastructure of NCHDC.

\vspace*{-0.25cm}
\bibliography{}% your bib database

\begin{thebibliography}{99}
\bibitem{Car}
C.~Carlet, {\em Boolean Functions for Cryptography and Coding Theory}, Cambridge University Press, Cambridge, 2021.
\bibitem{CarSol}
 C.~ Carlet and P.~ Sol\'{e}, "The weight spectrum of two families of Reed-Muller codes", Discrete Mathematics, {\bf 346(10)}, 113568, 2023.
\bibitem{CusSta}
Th.~W.~Cusick and P. St\u{a}nic\u{a}, {\em Cryptographic Boolean functions and Applications}, Academic Press, Amsterdam,\ldots, Tokyo, 2009.
\bibitem{GilLan}
V.~Gillot and Ph.~Langevin, "Classification of some cosets of the Reed-Muller code", Cryptogr. Commun. (2023),\\
available at https://doi.org/10.1007/s12095-023-00652-4. %\\https://hal-univ-tln.archives-ouverts.fr/hal-03834481, 2022.
\bibitem{AHu}
A.~Hulpke, "Computing with group orbits",
available at https://www.math.colostate.edu/
\bibitem{Hou}
X.~-D.~Hou, "$GL(m,2)$ acting on ${\mathcal R}(r,m)/{\mathcal R}(r-1,m)$", Discrete Mathematics, {\bf 149}, 99-122, 1996.
\bibitem{KasTok}
T.~Kasami and N.~Tokura, "On the weight structure of Reed-Muller codes", IEEE Trans. Info. Theory, 16 , 752-759, 1970.
\bibitem{KasTokAzu}
T.~Kasami, N.~Tokura, S.~Azumi, "On the weight enumeration of weights less than 2.5d of Reed-Muller codes", Information and Control, 30, 380-395, 1976.
\bibitem{Lan}
Ph.~Langevin, "Classification of Boolean quartic forms in eight variables",\\
available at https://langevin.univ-tln.fr/project/quartics/quartics.html, 2007.
\bibitem{Lan1}
Ph.~Langevin, "Classification of $RM(4,7)/RM(2,7)$",\\
available at https://langevin.univ-tln.fr/project/rm742/rm742.html, 2012.
\bibitem{LanLea}
Ph.~Langevin and G. Leander, "Classification of Boolean quartic forms in eight variables",
in {\em Boolean Functions in Cryptology and Information Security}, B.~Preneel and O.~A.~Logachev (Eds.), IOS Press, 139-147, 2008.
\bibitem{McWSlo}
F.~J.~MacWilliams and N.~J.~A.~Sloane, {\em The Theory of Error-Correcting Codes}, North-Holland Publishing Company,  Amsterdam, New York, Oxford, 1977.
%\bibitem{McE}
%R.~J.~McEliece, "On periodic sequences from GF (q)", J. Combin. Theory Ser. A, 10, 80-91, 1971.
\bibitem{DVS}
D.~V.~Sarwate, {\em Weight Enumeration of Reed-Muller Codes and Cosets}, Ph.D., Dep. Elec. Eng., Princeton Univ., Princeton, N.J., Sept. 1973, Advisors: E.~R.~Berlekamp and J.~D.~Ullman.
\bibitem{OEIS}
N.~J.~A. Sloane, "On-line Encyclopedia of Integer Sequences".
available at https://oeis.org/wiki/List of weight distributions
\bibitem{SloBer}
N.~J.~A. Sloane, E.~R.~Berlekamp, "Weight enumerator for second-order Reed-Muller codes", IEEE Trans. Info. Theory, 16, 745-751, 1970.
\bibitem{SugKasFuj}
Ts.~Sugita, T.~Kasami, and T.~Fujiwara, "The weight distribution of the third-order Reed-Muller code of length 512",  IEEE Trans. Info. Theory, 42, No. 5, 1622-1625, 1996.
\bibitem{SugTokKas}
M.~Sugino, Y.~Tokura and T.~Kasami, "Weight Distribution of (128,64) Reed-Muller Code", IEEE Trans. Info. Theory, 17, 627-628, 1971.
\bibitem{vTi}
H.~C.~A.~van Tilborg, "Weights in the third-order Reed-Muller codes", JPL Technical Report 32-1526, vol.IV, 86-92, 1971.
\end{thebibliography}

\vspace{0.6cm}
{\small

INSTITUTE OF MATHEMATICS AND INFORMATICS, BULGARIAN ACADEMY OF SCIENCES, 8 G BONCHEV STR., 1113
SOFIA, BULGARIA

\medskip
\emph{Email address}: miro@math.bas.bg

\vspace{0.5cm}
INSTITUTE OF MATHEMATICS AND INFORMATICS, BULGARIAN ACADEMY OF SCIENCES, 8 G BONCHEV STR., 1113
SOFIA, BULGARIA

\smallskip
\emph{Email address}: youri@math.bas.bg
}

\newpage
%\begin{landscape}
\begin{appendices}
\renewcommand{\thesection}{\appendixname~\Alph{section}} % or try \arabic{section}

\begin{landscape}
\section{}\label{appendixA}

\begin{table}[htbp!]\label{tab.1}
\centering
\caption{Sizes of partitions $\Delta(e)$ and $\Delta^{\prime}(e)$}
\label{table:original_to_merged_lengths}
\setlength{\tabcolsep}{2.0mm}{
\begin{tabular}{|l|r|r|}
       $e \in  {\mathcal E}(4,7)$: ANF's according to (\cite{Lan1})       & |$\Delta(e)|$ &  |$\Delta^{\prime}(e)$|  \\
\hline
  0                                   &      12  &      12   \\  %  e0
  4567                                &      63  &      52   \\  % e11
  1235+1345+1356+1456+2346+2356+2456  &     130  &     112   \\  %  e2
  2367+4567                           &     289  &     182   \\  % e10
  1237+4567                           &     480  &     306   \\  %  e7
  1257+1367+4567                      &     730  &     395   \\  %  e8
  1237+1247+1357+2367+4567            &     204  &     157   \\  %  e6
  1236+1257+1345+1467+2347+2456+3567  &    1098  &     675   \\  %  e3
  1236+1356+1567+2357+2467+2567+3456  &    1340  &     811   \\  %  e4
  1367+2345+2356+3456+4567            &    6449  &    2170   \\  %  e9
  1234+1237+1267+1567+2345+3456+4567  &   23988  &    3377   \\  %  e1
  1236+1367+1567+2345+3456+3457+3467  &   33660  &    4636   \\  %  e5
\end{tabular}}
\end{table}

\begin{table}[htbp!]
\centering
\caption{The matching between ${\mathcal E}^{\prime}(4,7)$ and ${\mathcal E}(4,7)$}
\label{table:equivalent_classes}
\setlength{\tabcolsep}{2.0mm}{
\begin{tabular}{|r|l|l|l|}
  Distribution of $n(e)$  &     ${\overline{\mathcal E}^{\prime}(4,7)}$     &  ${\overline{\mathcal E}(4,7)}$          &  Transition linear transform  \\
\hline
             3  &  0                        &  0                            &  [1000000 0100000 0010000 0001000 0000100 0000010 0000001] \\  %  e0
             2  &  123                      &  123                          &  [1000000 0100000 0010000 0001000 0000100 0000010 0000001] \\  % e11
            21  &  127+136+145              &  137+147+157+237+247+267+467  &  [0011001 0011110 0100110 1011000 1111010 1001100 0001100] \\  %  e2
            15  &  125+134                  &  123+145                      &  [1000000 0100000 0001000 0000100 0010000 0000010 0000001] \\  % e10
            89  &  126+345                  &  123+456                      &  [1000000 0100000 0001000 0000100 0000010 0010000 0000001] \\  %  e7
            56  &  126+135+234              &  123+245+346                  &  [0100000 0010000 0001000 0000010 0000100 1000000 0000001] \\  %  e8
            10  &  135+146+235+236+245      &  123+145+246+356+456          &  [1000000 0000010 0001000 0010000 0000100 0100000 0000001] \\  %  e6
             7  &  127+136+145+234          &  124+137+156+235+267+346+457  &  [0110001 1011001 0110011 0111010 1100101 0010111 1001011] \\  %  e3
           502  &  125+134+135+167+247+357  &  127+134+135+146+234+247+457  &  [0001000 0010000 0000001 0000100 0100000 0000010 1100110] \\  %  e4
             1  &  123+247+356              &  123+127+147+167+245          &  [0010000 0110011 1010000 0001110 0000001 0010011 0000100] \\  %  e9
             1  &  147+156+237+246+345      &  123+127+167+234+345+456+567  &  [0101010 1001010 1001001 1111111 0011000 0100010 1001011] \\  %  e1
           292  &  127+146+236+345          &  125+126+127+167+234+245+457  &  [0100111 0001110 0110110 1011000 0000010 0000100 0010110] \\  %  e5
\end{tabular}}
\end{table}
\end{landscape}

\newpage
\section{}\label{appendixC}

\begin{table}[htbp!]                                                    % https://tex.stackexchange.com/questions/8652/what-does-t-and-ht-mean
\centering
\caption{Weight Distribution of the (512,256,32) Reed-Muller code}
\label{table:weight_distrib}
\setlength{\tabcolsep}{2.0mm}{
\fontsize{3.4mm}{3.7mm}\selectfont
\begin{tabular}{rrr}
\toprule
\multicolumn{2}{c}{\textbf{Weight}} & \multicolumn{1}{c}{\textbf{Number of codewords}}          \\
\midrule
   0  &  512  &                                                                              1  \\
  32  &  480  &                                                                       52955952  \\
  48  &  464  &                                                                   919315326720  \\
  56  &  456  &                                                                271767121346560  \\
  60  &  452  &                                                                860689275027456  \\
  64  &  448  &                                                              89163020044002040  \\
  68  &  444  &                                                            1777323352931696640  \\
  72  &  440  &                                                           64959328938397057024  \\
  76  &  436  &                                                         2094952122987829002240  \\
  80  &  432  &                                                        86129855718211879936768  \\
  84  &  428  &                                                      3718387228743293604986880  \\
  88  &  424  &                                                    216407674400647746861465600  \\
  92  &  420  &                                                  15958945395035022932054114304  \\
  96  &  416  &                                                1570964763114053055495174389136  \\
 100  &  412  &                                              207755244457303752035637154283520  \\
 104  &  408  &                                            34164336816436357675455725024378880  \\
 108  &  404  &                                          5992987676360073735151889707696128000  \\
 112  &  400  &                                        983217921810034263357552475089021004288  \\
 116  &  396  &                                     140881159168600922710983130625456163782656  \\
 120  &  392  &                                   17178463264607761296016540993629780705771520  \\
 124  &  388  &                                 1770270551281316280504947079180771901717872640  \\
 128  &  384  &                               154198773988541804525321284585063483246993999900  \\
 132  &  380  &                             11380437366712812474455950864177326068447989202944  \\
 136  &  376  &                            713793445298874211607839796879716106185715280216064  \\
 140  &  372  &                          38161660034401312989486264769054124765959796671119360  \\
 144  &  368  &                        1744077996406613042017016863461234839306732612077058560  \\
 148  &  364  &                       68320936493023612641136928149296775084064365913214812160  \\
 152  &  360  &                     2299744204800465802453316637595783829108912802028206751744  \\
 156  &  356  &                    66674424868716978552789375387240003239187186349775851094016  \\
 160  &  352  &                  1668559700964160587350805664583122924498928358151715733007408  \\
 164  &  348  &                 36117082274027891545154187373048131661136552390031364702863360  \\
 168  &  344  &                677483598989547107793615101247739514269621184741356041461104640  \\
 172  &  340  &              11032441933713096201663286389373184730113421621201515757397082112  \\
 176  &  336  &             156225095497619813307679231937780861426835567156776476525084177664  \\
 180  &  332  &            1926667532217097161576702991776654344250440175688196887457279508480  \\
 184  &  328  &           20723534026876536792281002394151796205045793736436788802938336133120  \\
 188  &  324  &          194671442741837852939975553363771856234841259238404365556287065292800  \\
 192  &  320  &         1599044990181340998819270766161596605692512085057170791477694075282632  \\
 196  &  316  &        11498415685246302189888474222781442491860129957714864173250891967627264  \\
 200  &  312  &        72459467570743603819378812718772497540870770484626494838959726267809792  \\
 204  &  308  &       400549932263936554220342987258224499780564121712827465674395223861493760  \\
 208  &  304  &      1944071611978423909059426198144849863064608675044397429548995177751732480  \\
 212  &  300  &      8291211853278378544436157221213736835450108801042695204524353086973542400  \\
 216  &  296  &     31095502600701130763682713427899390240950550846409105550583369693522427904  \\
 220  &  292  &    102622652435510219354959437959897900434480615845926142166854426192158654464  \\
 224  &  288  &    298206281302110726623000750445450132512881810629607123478473554095237810960  \\
 228  &  284  &    763396919631666688676755106996803883003881847438728311891109384630797598720  \\
 232  &  280  &   1722452776176219896357452486934573175804665343735169479919087899582551687168  \\
 236  &  276  &   3426750460257305904470547641506642175867699465315478403354123631366508642304  \\
 240  &  272  &   6013163599489683999312799935491777179772724247998877953378442920501417933824  \\
 244  &  268  &   9309551320248854051332692772889245412495562988894547412532818045057116405760  \\
 248  &  264  &  12718986044129514620716674156341900030463015021774940408815989741288144568320  \\
 252  &  260  &  15336997499945305387056357527918950456934399969250231086077675815418680311808  \\
 \multicolumn{2}{c}{256} & \multicolumn{1}{r}{16324199909251682000435577287934368523097397692548071777837483832108326674502} \\
\bottomrule
\end{tabular}}
\end{table}

\end{appendices}

\end{document}